\documentclass[floats,showpacs, nofootinbib, preprint,
eqsecnum]{revtex4-1}

\usepackage{color,graphicx}
\usepackage{amsfonts}
\usepackage{amssymb}
\begin{document}

\title{Retrolensing by a charged black hole}

\author{Naoki Tsukamoto}\email{tsukamoto@rikkyo.ac.jp}
\author{Yungui Gong}\email{yggong@mail.hust.edu.cn}

\affiliation{
School of Physics, Huazhong University of Science and Technology, Wuhan 430074, China
}

\begin{abstract}
Compact objects with a light sphere such as black holes and wormholes can reflect light rays like a mirror.
This gravitational lensing phenomenon is called retrolensing and it is an interesting tool to survey dark and compact objects
with a light sphere near the solar system.
In this paper, we calculate the deflection angle analytically in the strong deflection limit
in the Reissner-Nordstr\"{o}m spacetime without Taylor expanding it in the power of the electric charge.
Using the obtained deflection angle in the strong deflection limit,
we investigate the retrolensing light curves and the separation of double images by the light sphere of Reissner-Nordstr\"{o}m black holes.
\end{abstract}

\maketitle

\section{Introduction}
Gravitational lensing is an important phenomenon to survey dark and compact objects and to test gravitational theories
from the scale of the solar system to the cosmological size.
Gravitational lensing was investigated mainly under the quasi-Newtonian or weak-field approximation~\cite{Schneider_Ehlers_Falco_1992,Petters_Levine_Wambsganss_2001,Schneider_Kochanek_Wambsganss_2006,Bartelmann_2010},
but gravitational lensing beyond the quasi-Newtonian approximation was studied also~\cite{Perlick_2004_Living_Rev,Bozza_2010}.
Since Darwin pointed out that faint images near the circular orbit of a light ray
called light sphere or photon sphere~\cite{Claudel:2000yi,Hasse_Perlick_2002} appear in the Schwarzschild spacetime in 1959~\cite{Darwin_1959},
the study of the gravitational lensing phenomena was revived several times~\cite{Atkinson_1965,Luminet_1979,Ohanian_1987,Nemiroff_1993%
,Frittelli_Kling_Newman_2000,Virbhadra_Ellis_2000,Bozza_Capozziello_Iovane_Scarpetta_2001,Bozza_2002,Perlick_2004_Phys_Rev_D%
,Iyer:2006cn,Virbhadra_2009,Bozza_Sereno_2006,Bozza:2007gt,Bozza:2008ev,Ishihara:2016sfv}.
The gravitational lensing by the light sphere in
various black hole spacetimes~\cite{Bozza:2002af,Bozza:2005tg,Bozza:2006nm,Saida:2016kpk,Chen:2016hil,Bozza_2002,Eiroa:2002mk,Eiroa:2004gh,Whisker:2004gq,Eiroa:2012fb,Bhadra:2003zs,Eiroa:2005ag,Mukherjee:2006ru,Gyulchev:2006zg,Chen:2009eu,Liu:2010wh,Eiroa:2010wm,Ding:2010dc,Chen:2011ef,Wei:2011nj,AzregAinou:2012xv,Gyulchev:2012ty,Eiroa:2013nra,Tsukamoto:2014dta,Wei:2014dka,Eiroa:2014mca,Sahu:2015dea,Sotani:2015ewa,Zhao:2016kft,Chakraborty:2016lxo},
naked singularity spacetimes~\cite{Virbhadra_Keeton_2008,Virbhadra_Ellis_2002,Dey_Sen_2008,Bozza_2002},
boson star spacetimes~\cite{Horvat:2013plm,Cunha:2015yba,Cunha:2016bjh},
and wormhole spacetimes~\cite{Nandi_Zhang_Zakharov_2006,Dey_Sen_2008,Perlick_2004_Phys_Rev_D,Tsukamoto_Harada_Yajima_2012,Tsukamoto:2016qro,Nandi:2016ccg,Tsukamoto:2016zdu},
was also investigated.

Recently, LIGO reported a gravitational-wave event GW150914~\cite{Abbott:2016blz} and showed the existence
of heavy stellar-mass black holes with the mass $M\geq  25M_{\odot}$ in nature~\cite{TheLIGOScientific:2016htt}.
Black holes become an important subject in astronomy and astrophysics,
and gravitational lensing by black holes will be an important tool to study dim and isolated black holes.

Holz and Wheeler~\cite{Holz:2002uf} proposed to survey retrolensing caused by sun lights which were reflected by the light sphere of stellar-mass black holes
near the solar system in 2002.
A black hole in the Galactic center as a retrolens~\cite{DePaolis:2003ad,Eiroa:2003jf,Bozza:2004kq},
wormholes as retrolenses~\cite{Tsukamoto:2016qro},
and the effects by the rotation~\cite{DePaolis:2004xe,Bozza:2004kq} and the electric charge~\cite{Eiroa:2003jf} of a black hole on the magnification
were also studied.
In Ref.~\cite{Bozza_2002}, Bozza considered gravitational lensing effects of light rays passed near the light sphere
in a general static and spherical symmetric spacetime and provided a formula for the deflection angle in the strong deflection limit.
The deflection angle in the strong deflection limit describes a fundamental feature of a light sphere.
The relation between the deflection angle in the strong deflection limit and the quasinormal modes~\cite{Stefanov:2010xz,Wei:2013mda},
and the high-energy absorption cross section~\cite{Wei:2011zw} were also discussed.

Eiroa and Torres~\cite{Eiroa:2003jf} investigated retrolensing of light rays with deflection angle in the strong deflection limit
in the general static and spherical symmetric spacetime.
They discussed the image angles and magnifications of images of light rays
emitted by an extended source and deflected by a light sphere of a charged black hole
when the source, observer, and black hole are almost aligned.
Bozza and Mancini~\cite{Bozza:2004kq} considered a general treatment of light rays with a deflection angle in the strong deflection limit
without the almost-aligned assumption in the general static and spherical symmetric spacetime and they used retrolensing as an example.

Considering the difference of observables between Schwarzschild black holes and other black holes
is important to check whether an observed non-rotating black hole can be really described by the Schwarzschild black hole.
The Reissner-Nordstr\"{o}m black hole solution is one of the useful and simple black hole solutions for the purpose
even though black holes in nature would be almost electrically neutralized.
The deflection angle in the strong deflection limit in the Reissner-Nordstr\"{o}m black hole spacetime
was obtained numerically first in the pioneering work by Eiroa \textit{et al.}  in Ref.~\cite{Eiroa:2002mk}.
Bozza investigated the deflection angle in the strong deflection limit in the Reissner-Nordstr\"{o}m black hole spacetime
and pointed out an integral in the deflection angle cannot be calculated analytically in Ref.~\cite{Bozza_2002}.
The integral was calculated numerically
and approximately by Taylor expanding the integral in the power of the electric charge~\cite{Bozza_2002}.

In this paper, we revisit the analysis made by Bozza~\cite{Bozza_2002} in the strong deflection limit
and obtain the analytical result for the deflection angle in the strong deflection limit in the Reissner-Nordstr\"{o}m black hole spacetime,
and we also consider the retrolensing by the Reissner-Nordstr\"{o}m black hole.
We consider the details of the effect of the electric charge on retrolensing light curves and the separation of retrolensing double images
by a stellar mass black hole near the solar system which might be measured with future observations.
We discuss the estimation of the mass and charge of the black hole, and the distance between the black hole and the observer
from the shape and magnitude of the retrolensing light curve
and the image separation of retrolensing double images.

This paper is organized as follows.
In Sec.~II we investigate the deflection angle in the strong deflection limit in the Reissner-Nordstr\"{o}m spacetime
and obtain the analytical formula without Taylor expanding it in the power of the electric charge.
In Sec.~III we review retrolensing and
investigate the effect of electric charges on retrolensing light curves and double images by a retrolens in the Reissner-Nordstr\"{o}m spacetime.
In Sec.~IV we conclude the paper.
In this paper we use the unit in which the light speed and Newton's constant are unity.

\section{Deflection angle in the strong deflection limit in the Reissner-Nordstr\"{o}m spacetime}
In this section, we calculate the deflection angle $\alpha$ of a light ray in the strong deflection limit in the Reissner-Nordstr\"{o}m spacetime with the following form:
\begin{equation}\label{eq:deflection0}
\alpha(b)= -\bar{a}\log \left( \frac{b}{b_{c}}-1 \right) +\bar{b}+O((b-b_{c})\log (b-b_{c})),
\end{equation}
where $b$ is the impact parameter of the light ray, $b_{c}$ is the critical impact parameter of the light ray,
$\bar{a}$ is a positive function of $Q/M$ which is the ratio of the electric charge $Q$
to the Arnowitt-Deser-Misner mass $M$ of a Reissner-Nordstr\"{o}m black hole,
and $\bar{b}$ is a negative function of $Q/M$
\footnote{In~\cite{Bozza_2002}, the order of magnitude of an error term in the deflection angle in the strong deflection limit is estimated as $O(b-b_{c})$.
Recently, however, Tsukamoto pointed out that the order of magnitude of the error term should be read as $O((b-b_{c})\log (b-b_{c}))$.
See Ref.~\cite{Tsukamoto:2016qro} for more details.}.
The line element of the Reissner-Nordstr\"{o}m spacetime is given by
\begin{equation}
\label{rnmetriceq}
ds^{2}
=-\frac{\Delta(r)}{r^{2}}dt^{2} +\frac{r^{2}}{\Delta(r)}dr^{2}+r^{2}(d\theta^{2}+\sin^{2}\theta d\phi^{2}),
\end{equation}
where $\Delta(r)=r^{2}-2Mr+Q^{2}$.
Because the spacetime is a static and spherical symmetric one,
there are time translational and  axial Killing vectors $t^{\mu}\partial_{\mu}=\partial_{t}$ and $\phi^{\mu}\partial_{\mu}=\partial_{\phi}$,
respectively.
The Reissner-Nordstr\"{o}m spacetime is a black hole spacetime with an event horizon at $r=r_{H}$, where
\begin{equation}\label{eq:event_horizon}
r_{H}\equiv M+\sqrt{M^{2}-Q^{2}}
\end{equation}
for $Q\leq M$
while it is a naked singularity spacetime
for $M < Q$.
We concentrate on the black hole spacetime, i.e., $Q\leq M$.
Please note that $r=r_{H}$ is the largest positive solution of the equation $\Delta(r)=0$.
The radius of a light sphere $r_{m}$ is the largest positive solution of the equation~\cite{Hasse_Perlick_2002}
\begin{equation}
\label{eq:2.5}
\frac{g_{\theta\theta}'(r)}{g_{\theta\theta}(r)}-\frac{g_{tt}'(r)}{g_{tt}(r)}=0,
\end{equation}
where $'$ is the differentiation with respect to the radial coordinate $r$.
From Eqs. (\ref{rnmetriceq}) and (\ref{eq:2.5}), we get
\begin{equation}\label{eq:rm2}
r_{m}^{2}-3Mr_{m}+2Q^{2}=0.
\end{equation}
Thus, the light sphere exists at $r=r_{m}$, where
\begin{equation}\label{eq:rm1}
r_{m}=\frac{3M+\sqrt{9M^{2}-8Q^{2}}}{2}.
\end{equation}

The trajectory of a light ray is described by
\begin{equation}\label{eq:trajectory1}
k^{\mu}k_{\mu}=0,
\end{equation}
where $k^{\mu}\equiv \dot{x}^{\mu}$ is the wave number of the light ray
and the dot denotes the differentiation with respect to an affine parameter parametrizing the null geodesic.
The conserved energy $E$ and angular momentum $L$ of the photon given by
\begin{equation}
E\equiv -g_{\mu\nu}t^{\mu}k^{\nu}=\frac{\Delta(r)}{r^{2}}\dot{t},
\end{equation}
and
\begin{equation}
L\equiv g_{\mu\nu}\phi^{\mu}k^{\nu}=r^{2}\dot{\phi},
\end{equation}
respectively, are constant along the null geodesic.
We assume that $E$ is positive.
The impact parameter $b$ of a light ray is defined as
\begin{equation}
b\equiv \frac{L}{E}=\frac{r^{4}\dot{\phi}}{\Delta(r)\dot{t}}.
\end{equation}
We assume $L$ and $b$ are positive without the loss of generality as long as we consider one light ray.
We also assume $\theta=\pi/2$ without the loss of generality.

From Eq.~(\ref{eq:trajectory1}), we get the trajectory equation in the Reissner-Nordstr\"{o}m spacetime
\begin{equation}\label{eq:trajectory2}
-\frac{\Delta}{r^{2}}\dot{t}^{2}+\frac{r^{2}}{\Delta}\dot{r}^{2}+r^{2}\dot{\phi}^{2}=0,
\end{equation}
or
\begin{equation}\label{eq:trajectory3}
\dot{r}^{2}=V(r),
\end{equation}
where the effective potential $V(r)$ for the motion of the photon is
\begin{equation}\label{eq:V}
V(r)\equiv E^{2}-\frac{\Delta(r)}{r^{4}}L^{2}.
\end{equation}
The motion of the photon is permitted in a region where $V(r)$ is non-negative.
Since the effective potential $V(r)$ in the limit $r\rightarrow \infty$ becomes
\begin{equation}\label{eq:V1}
V(r)\rightarrow E^{2} >0,
\end{equation}
a photon can be at the spatial infinity $r \rightarrow \infty$.
We consider that a photon coming from infinity, is reflected by a black hole at the closest distance $r=r_{0}$, and goes to infinity.
Since $\dot{r}$ vanishes at the closest distance $r=r_{0}$, from Eqs.~(\ref{eq:trajectory3}) and~(\ref{eq:V}),
we obtain a relation between the impact parameter $b$ and the closest distance $r_{0}$ as
\begin{equation}\label{eq:b1}
b(r_{0})=\frac{r_{0}^{2}}{\sqrt{\Delta_{0}}},
\end{equation}
where $\Delta_{0}\equiv \Delta(r_{0})$.
Hereafter subscript $0$ denotes the quantity at the closest distance $r=r_{0}$.

We define the critical impact parameter $b_{c}$ as
\begin{equation}\label{eq:bc1}
b_{c}(r_{m})
\equiv \lim_{r_{0}\rightarrow r_{m}} b(r_{0})
=\lim_{r_{0}\rightarrow r_{m}} \frac{r_{0}^{2}}{\sqrt{\Delta (r_{0})}}.
\end{equation}
The limit $r_{0}\rightarrow r_{m}$ or $b\rightarrow b_{c}$ is referred to as the strong deflection limit.
The impact parameter $b(r_{0})$ can be expanded in the power of $r_{0}-r_{m}$ as
\begin{eqnarray}\label{eq:b2}
b(r_{0})
&=&b_{c}(r_{m})+\frac{3Mr_{m}-4Q^{2}}{2(Mr_{m}-Q^{2})^{\frac{3}{2}}} (r_{0}-r_{m})^{2} \nonumber\\
&&+O\left((r_{0}-r_{m})^{3}\right).
\end{eqnarray}
From the derivative of the effective potential $V(r)$ with respect to the radial coordinate $r$,
\begin{equation}
V'(r)=\frac{2L(r^{2}-3Mr+2Q^{2})}{r^{5}},
\end{equation}
and Eqs.~(\ref{eq:rm2}) and (\ref{eq:trajectory3}), we obtain
\begin{equation}
\lim_{r_{0}\rightarrow r_{m}}V(r_{0})=0
\end{equation}
and
\begin{equation}
\lim_{r_{0}\rightarrow r_{m}}V'(r_{0})=0.
\end{equation}
Thus, in the strong deflection limit $r_{0}\rightarrow r_{m}$ or $b\rightarrow b_{c}$,
a photon with the impact parameter $b\rightarrow b_{c}$ almost stops in the radial direction near outside the light sphere at $r=r_{m}$
and the orbit of the photon winds around the light sphere.

The trajectory equation~(\ref{eq:trajectory2}) can be rewritten as
\begin{equation}\label{eq:trajectory4}
\left( \frac{dr}{d\phi} \right)^{2}=r^{4} \left( \frac{1}{b^{2}}-\frac{\Delta}{r^{4}} \right)
\end{equation}
and the deflection angle $\alpha(r_{0})$ of the light ray is obtained as
\begin{equation}\label{eq:exact_declection}
\alpha = I(r_{0})-\pi,
\end{equation}
where
\begin{equation}
I(r_{0}) \equiv 2\int^{\infty}_{r_{0}} \frac{dr}{r^{2}\sqrt{\frac{1}{b^{2}}-\frac{\Delta}{r^{4}}}}.
\end{equation}

By introducing a variable $z$ defined as~\footnote{In Ref.~\cite{Bozza_2002}, Bozza introduced a variable $z_{[17]}$ defined by
\begin{equation}
z_{[17]}\equiv \frac{-g_{tt}(r)+g_{tt}(r_{0})}{1+g_{tt}(r_{0})}
\end{equation}
in our notation.
See Eqs. (10) and (11) in~\cite{Bozza_2002}.
The difference between $z$ and $z_{[17]}$ is discussed in Sec.~IV in this paper.}
\begin{equation}\label{eq:z1}
z\equiv 1-\frac{r_{0}}{r},
\end{equation}
$I(r_{0})$ can be rewritten as
\begin{equation}
I(r_{0})=\int^{1}_{0}f(z,r_{0})dz,
\end{equation}
where
\begin{equation}
f(z,r_{0})\equiv \frac{2r_{0}}{\sqrt{c_{1}(r_{0})z+c_{2}(r_{0})z^{2}+c_{3}(r_{0})z^{3}+c_{4}(r_{0})z^{4}}},
\end{equation}
\begin{equation}
c_{1}(r_{0})\equiv 2(r_{0}^{2}-3Mr_{0}+2Q^{2}),
\end{equation}
\begin{equation}
c_{2}(r_{0})\equiv -r_{0}^{2}+6Mr_{0}-6Q^{2},
\end{equation}
\begin{equation}
c_{3}(r_{0})\equiv -2Mr_{0}+4Q^{2},
\end{equation}
and
\begin{equation}
c_{4}(r_{0})\equiv -Q^{2}.
\end{equation}

Since $c_{1}(r_{0})$ and $c_{2}(r_{0})$ in the strong deflection limit $r_{0}\rightarrow r_{m}$ become
\begin{equation}
c_{1}(r_{0}) \rightarrow 0
\end{equation}
and
\begin{equation}
c_{2}(r_{0}) \rightarrow 3Mr_{m}-4Q^{2},
\end{equation}
respectively,
the order of the divergence of $f(z,r_{0})$ is $z^{-1}$.
We separate $I(r_{0})$ into two parts, i.e., a divergent part $I_{D}(r_{0})$ and a regular part $I_{R}(r_{0})$:
$I(r_{0})=I_{D}(r_{0})+I_{R}(r_{0})$.
The divergent part $I_{D}(r_{0})$ is defined as
\begin{equation}
I_{D}(r_{0})\equiv \int^{1}_{0}f_{D}(z,r_{0})dz,
\end{equation}
where
\begin{equation}
f_{D}(z,r_{0})\equiv \frac{2r_{0}}{\sqrt{c_{1}(r_{0})z+c_{2}(r_{0})z^{2}}},
\end{equation}
and the result is
\begin{eqnarray}
I_{D}(r_{0})=&&\frac{4r_{0}}{\sqrt{-r_{0}^{2}+6Mr_{0}-6Q^{2}}}  \nonumber\\
&&\times \log \frac{\sqrt{-r_{0}^{2}+6Mr_{0}-6Q^{2}}+\sqrt{r_{0}^{2}-2Q^{2}}}{\sqrt{2(r_{0}^{2}-3Mr_{0}+2Q^{2})}}. \nonumber\\
\end{eqnarray}
Using Eq.~(\ref{eq:b2}), $I_{D}$ in the strong deflection limit $r_{0}\rightarrow r_{m}$ or $b \rightarrow b_{c}$ is expressed as
\begin{eqnarray}
I_{D}(b)
=&& -\bar{a} \log \left( \frac{b}{b_{c}}-1 \right) +\bar{a} \log \frac{2(3Mr_{m}-4Q^{2})}{Mr_{m}-Q^{2}} \nonumber\\
&& +O((b-b_{c})\log (b-b_{c})),
\end{eqnarray}
where $\bar{a}$ is given by
\begin{equation}
\bar{a}=\frac{r_{m}}{\sqrt{3Mr_{m}-4Q^{2}}}.
\end{equation}
We will see that in the strong deflection limit $\bar{a}$ is the function appearing in Eq.~(\ref{eq:deflection0}) later.

The regular part $I_{R}$ is defined as
\begin{equation}
I_{R}(r_{0})\equiv \int^{1}_{0}f_{R}(z,r_{0})dz,
\end{equation}
where
\begin{equation}
f_{R}(z,r_{0}) \equiv f(z,r_{0})-f_{D}(z,r_{0}).
\end{equation}
Since we are interested in the deflection angle in the strong deflection limit $r_{0}\rightarrow r_{m}$,
we consider
\begin{eqnarray}
\lim_{r_{0} \rightarrow r_{m}}f_{R}(z,r_{0})
&=&\frac{2r_{m}}{z\sqrt{c_{2}(r_{m})+c_{3}(r_{m})z+c_{4}(r_{m})z^{2}}}\nonumber\\
&&-\frac{2r_{m}}{z\sqrt{c_{2}(r_{m})}}.
\end{eqnarray}
In the strong deflection limit $r_{0}\rightarrow r_{m}$ or $b \rightarrow b_{c}$, we obtain the analytical expression,
\begin{eqnarray}\label{eq:IR}
I_{R}(b)
&=& \bar{a}
\log \left[ \frac{4(3Mr_{m}-4Q^{2})^{2}}{M^{2}r_{m}^{2}(Mr_{m}-Q^{2})} \right. \nonumber\\
&&\left. \times \left(2\sqrt{Mr_{m}-Q^{2}}-\sqrt{3Mr_{m}-4Q^{2}}\right)^{2} \right] \nonumber\\
&& +O((b-b_{c})\log (b-b_{c})).
\end{eqnarray}

\begin{figure}[htbp]
\begin{center}
\includegraphics[width=70mm]{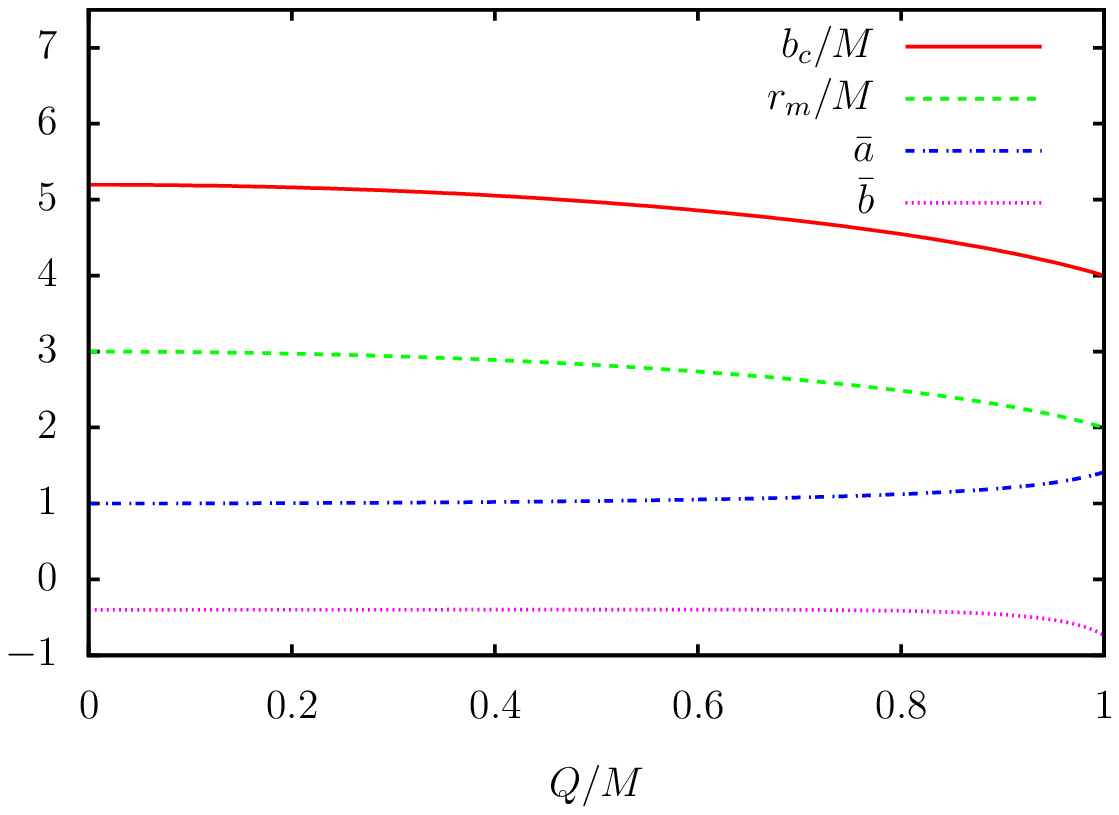}
\end{center}
\caption{The quantities $b_{c}/M$, $r_{m}/M$, $\bar{a}$, and $\bar{b}$ in the Reissner-Nordstr\"{o}m
spacetime as the functions of $Q/M$. The (red) solid, (green) dashed,
(blue) dash-dotted, and (purple) dotted curves denote $b_{c}/M$, $r_{m}/M$, $\bar{a}$, and $\bar{b}$, respectively.}
\label{parameter}
\end{figure}

Thus, the deflection angle $\alpha(b)$ in the strong deflection limit $b\rightarrow b_{c}$ is given by
\begin{equation}\label{eq:deflection1}
\alpha(b)= -\bar{a}\log \left( \frac{b}{b_{c}}-1 \right) +\bar{b}+O((b-b_{c})\log (b-b_{c})),
\end{equation}
where $\bar{a}$ and $\bar{b}$ are obtained as
\begin{equation}\label{eq:abar1}
\bar{a}=\frac{r_{m}}{\sqrt{3Mr_{m}-4Q^{2}}}
\end{equation}
and
\begin{eqnarray}\label{eq:bbar}
\bar{b}
&=&\bar{a} \log \left[ \frac{8(3Mr_{m}-4Q^{2})^{3}}{M^{2}r_{m}^{2}(Mr_{m}-Q^{2})^{2}} \right. \nonumber\\
&&\left. \times \left(2\sqrt{Mr_{m}-Q^{2}}-\sqrt{3Mr_{m}-4Q^{2}} \right)^{2} \right] -\pi, \nonumber\\
\end{eqnarray}
respectively.
Figure~\ref{parameter} shows $b_{c}/M$, $r_{m}/M$, $\bar{a}$ and $\bar{b}$ as the functions of $Q/M$.
When the black hole carries no charge ($Q=0$), we obtain $b_{c}=3\sqrt{3}M$, $\bar{a}=1$, and $\bar{b}=\log\left[216(7-4\sqrt{3})\right]-\pi$,
and we recover the well-known result in Refs.~\cite{Darwin_1959,Bozza_2002} for the Schwarzschild spacetime.
When the black hole has the maximal charge ($Q=M$), we obtain $b_{c}=4M$, $\bar{a}=\sqrt{2}$, and $\bar{b}=2\sqrt{2}\log\left[4(2-\sqrt{2})\right]-\pi$.

\subsection{Comparison with the result in Eiroa \textit{et al}. 2002~\cite{Eiroa:2002mk}}

Eiroa \textit{et al.}~\cite{Eiroa:2002mk} numerically obtained the deflection angle in the strong deflection
limit $r_{0}\rightarrow r_{m}$ in the Reissner-Nordstr\"{o}m spacetime.
In our notation, they obtained the values of $\mathcal{A}$ and $\mathcal{B}$ numerically by assuming the relation
\begin{equation}
\lim_{r_{0}\rightarrow r_{m}} \left( \alpha + \mathcal{A} \log \left[ \frac{\mathcal{B}(r_{0}-r_{m})}{2M} \right] +\pi \right)=0,
\end{equation}
where $\alpha$ is given by Eq.~(\ref{eq:exact_declection}).~\footnote{
In Ref.~\cite{Eiroa:2002mk}, $\mathcal{A}$ and $\mathcal{B}$ are denoted by $A$ and $B$, respectively.
}

We analytically derive $\mathcal{A}$ and $\mathcal{B}$ as
\begin{equation}\label{eq:matecalA}
\mathcal{A}\equiv 2\bar{a}
\end{equation}
and
\begin{equation}\label{eq:matecalB}
\mathcal{B}\equiv \frac{M}{\bar{a}}\sqrt{\frac{2}{Mr_{m}-Q^{2}}}\exp \left( -\frac{\bar{b}+\pi}{2\bar{a}} \right),
\end{equation}
respectively, where
$r_{m}$, $\bar{a}$, and $\bar{b}$ are given by Eqs.~(\ref{eq:rm1}), (\ref{eq:abar1}), and (\ref{eq:bbar}), respectively.
Table~\ref{table:I} shows $\mathcal{A}$ and $\mathcal{B}$ as functions of $Q/M$ obtained in this paper and in~\cite{Eiroa:2002mk}.
Note that $\mathcal{A}$ and $\mathcal{B}$ calculated analytically from Eqs. (\ref{eq:matecalA}) and (\ref{eq:matecalB}), respectively,
reproduce the numerical values given in Table I of Ref.~\cite{Eiroa:2002mk} with high precision.
\begin{table*}[hbtp]
 \caption{$\mathcal{A}$ and $\mathcal{B}$ of the deflection angle in the strong deflection limit in the Reissner-Nordstr\"{o}m spacetime.
 The numerical values of $\mathcal{A}$ and $\mathcal{B}$ are taken from Table~I in Ref.~\cite{Eiroa:2002mk}.
The same table was also shown in Ref.~\cite{Tsukamoto:2016jzh}.}
 \label{table:I}
\begin{center}
\begin{tabular}{c c c c c c c} \hline
$Q/M$ &$0$ &$0.1$ &$0.25$ &$0.5$ &$0.75$ &$1$ \\ \hline
$\mathcal{A}$ &$2.00000$  &$2.00224$  &$2.01444$  &$2.06586$  &$2.19737$  &$2.82843$ \\ \hline
$\mathcal{A}$ in~\cite{Eiroa:2002mk} &$2.00000$  &$2.00224$  &$2.01444$  &$2.06586$  &$2.19737$  &$2.82843$ \\ \hline
$\mathcal{B}$ &$0.207336$ &$0.207977$ &$0.211467$ &$0.225996$ &$0.262083$ &$0.426777$ \\ \hline
$\mathcal{B}$ in~\cite{Eiroa:2002mk} &$0.207338$ &$0.207979$ &$0.21147$  &$0.225997$ &$0.262085$ &$0.426782$ \\ \hline
\end{tabular}
\end{center}
\end{table*}

\subsection{Comparison with the result in Bozza 2002~\cite{Bozza_2002}}
In~\cite{Bozza_2002}, $\bar{a}$ in the Reissner-Nordstr\"{o}m spacetime was obtained in our notation as
\begin{equation}\label{eq:abar2}
\bar{a}=\frac{r_{m}\sqrt{Mr_{m}-Q^{2}}}{\sqrt{M(6M-r_{m})r^{2}_{m}-9r_{m}MQ^{2}+4Q^{4}}}.
\end{equation}
Using Eq.~(\ref{eq:rm2}),
one shows that Eq.~(\ref{eq:abar2}) is the same as Eq.~(\ref{eq:abar1}).~\footnote{Note that the electrical charge $q$ in Ref.~\cite{Bozza_2002}
denotes $q=2Q$.}

Bozza combined numerical and analytical calculations to obtain the numerical value of $\bar{b}$ in Ref.~\cite{Bozza_2002}.
Table~\ref{table:II} compares the value of $\bar{b}$ from Eq. (\ref{eq:bbar}) with that in~\cite{Bozza_2002} as functions of $Q/M$.
Note that the analytical expression (\ref{eq:bbar})
reproduces the numerical results given in Table I of Ref.~\cite{Bozza_2002} with high precision.
We also find that $\bar{b}$ does not decrease monotonically.
\begin{table}[hbtp]
 \caption{$\bar{b}$ in the deflection angle in the strong deflection limit.
The numerical values of $\bar{b}$ are taken from Table~I in Ref.~\cite{Bozza_2002}.}
 \label{table:II}
\begin{center}
\begin{tabular}{c c c c c c} \hline
$Q/M$ &$0$ &$0.2$ &$0.4$ &$0.6$ &$0.8$ \\ \hline
$\bar{b}$                      &$-0.4002$      &$-0.39935$    &$-0.3972$     &$-0.3965$     &$-0.4136$     \\ \hline
$\bar{b}$ in~\cite{Bozza_2002} &$-0.4002$      &$-0.3993$     &$-0.3972$     &$-0.3965$     &$-0.4136$     \\ \hline
\end{tabular}
\end{center}
\end{table}

\section{Retrolensing in the Reissner-Nordstr\"{o}m spacetime}
In this section, we review retrolensing following Ref.~~\cite{Bozza:2004kq}
and then we investigate retrolensing in the Reissner-Nordstr\"{o}m spacetime.

\subsection {Lens equation}
\begin{figure}[htbp]
\begin{center}
\includegraphics[width=70mm]{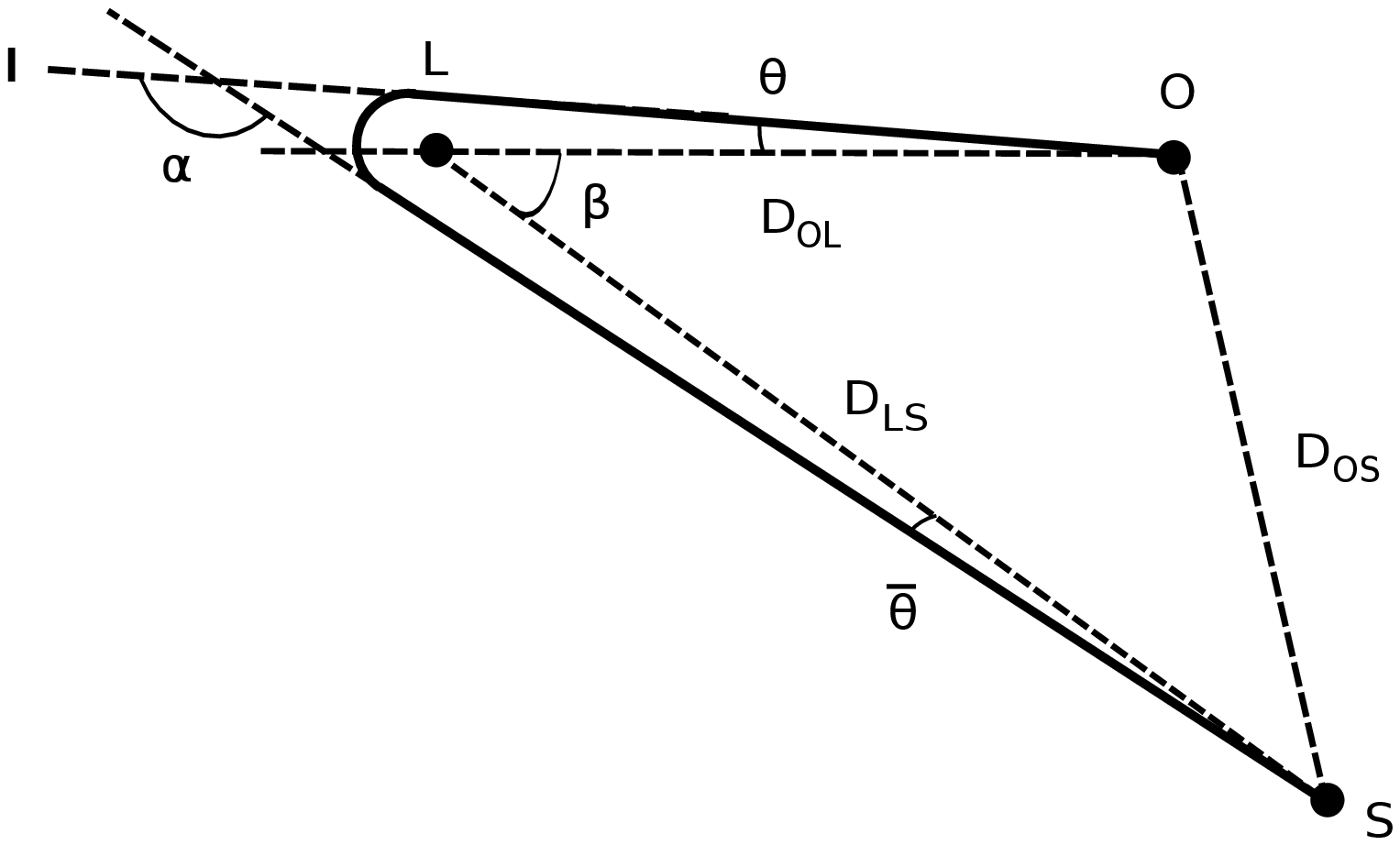}
\end{center}
\caption{Lens configuration.
The ray of the sun $S$ is reflected by a light sphere $L$ of a black hole with a deflection angle $\alpha$ and reaches the observer $O$.
The observer sees an image $I$ with an image angle $\theta$.
$\beta$ is the source angle $\angle OLS$. $\bar{\theta}$ is the angle between the line $LS$ and the light ray at $S$.}
\label{Lens_Configuration}
\end{figure}

We consider that the ray of the sun $S$ is reflected by the light sphere $L$ of a black hole
with a deflection angle $\alpha$ and reaches the observer $O$.
The observer sees an image $I$ with an image angle $\theta$.
The lens configuration is shown in Fig.~\ref{Lens_Configuration}.

We solve the lens equation proposed by Ohanian~\cite{Ohanian_1987,Bozza:2004kq,Bozza:2008ev},
\begin{equation}\label{eq:Lens1}
\beta=\pi-\alpha(\theta)+\theta+\bar{\theta},
\end{equation}
where $\beta$ is the source angle $\angle OLS$ defined in the range $0 \leq \beta \leq \pi$
and $\bar{\theta}$ is the angle between the line $LS$ and the light ray at $S$.
We assume that the black hole, the observer, and the sun are almost aligned in this order. Under the assumption, we obtain
\begin{equation}
\beta \sim 0
\end{equation}
and
\begin{equation}
D_{LS}=D_{OL}+D_{OS},
\end{equation}
where $D_{LS}$, $D_{OL}$, and $D_{OS}$ are the distances
between the black hole and the sun, between the black hole and the observer, and between the observer and the sun, respectively.
We concentrate on a case where the impact parameter $b$ is positive
and we assume that the radius of the light sphere is much smaller than $D_{OL}$ and $D_{LS}$,
i.e., $b_{c}\ll D_{OL}$ and $b_{c}\ll D_{LS}$.

\subsection{Image angle and magnification}
Neglecting the small terms $\theta=b/D_{OL}$ and $\bar{\theta}=b/D_{LS}$ in the Ohanian lens equation~(\ref{eq:Lens1})
and inserting the deflection angle $\alpha(b)$ obtained in Eq.~(\ref{eq:deflection1}) under the strong deflection limit
and $b=\theta D_{OL}$ into the Ohanian lens equation~(\ref{eq:Lens1}),
we obtain the positive solution $\theta=\theta_{+}(\beta)$ of the Ohanian lens equation as~\cite{Bozza:2004kq}
\begin{equation}\label{eq:theta+1}
\theta_{+}(\beta)\equiv \theta_{m} \left[ 1+\exp\left( \frac{\bar{b}-\pi+\beta}{\bar{a}} \right) \right],
\end{equation}
where $\theta_{m}\equiv b_{c}/D_{OL}$ is the image angle of the light sphere of the black hole.

The magnification $\mu_{+}$ of the image with the image angles $\theta_{+}$ is given by~\cite{Bozza:2004kq}
\begin{equation}\label{eq:magnification1}
\mu_{+}(\beta)=-\frac{D_{OS}^{2}}{D_{LS}^{2}} s(\beta)\theta_{+}\frac{d\theta_{+}}{d\beta},
\end{equation}
where
\begin{equation}
s(\beta)=\frac{1}{\beta}
\end{equation}
for a point source.
We assume that the sun can be described by a uniform-luminous disk with a finite size on the observer's sky~\cite{Witt:1994,Nemiroff:1994uz,Alcock:1997fi}.
In an uniform-luminous and finite-size source case, $s(\beta)$ is given by an integral over the disk on the source plane,
\begin{equation}
s(\beta)
= \frac{1}{\pi \beta_{s}^{2}} \int_{Disk} d\beta' d\phi,
\end{equation}
where $\beta'$ is a nondimensionalized radial coordinate divided by $D_{LS}$ on the source plane,
$\phi$ is the azimuthal coordinate around the origin on the source plane,
$\beta_{s}\equiv R_{s}/D_{LS}$, and $R_{s}$ is the radius of the sun.
We fix an intersection point of the axis $\beta=0$ and the source plane as the origin of the coordinates on the source plane
and then $s(\beta)$ is expressed as
\begin{eqnarray}
s(\beta)
&=& \frac{2}{\pi \beta_{s}^{2}} \left[ \pi(\beta_{s}-\beta) \right. \nonumber\\
&&\left.+\int^{\beta+\beta_{s}}_{-\beta+\beta_{s}} \arccos  \frac{\beta^{2}+\beta'^{2}-\beta^{2}_{s}}{2\beta \beta'} d\beta' \right]
\end{eqnarray}
for $\beta \leq \beta_{s}$ and
\begin{eqnarray}
s(\beta)
=\frac{2}{\pi \beta_{s}^{2}} \int^{\beta+\beta_{s}}_{\beta-\beta_{s}} \arccos \frac{\beta^{2}+\beta'^{2}-\beta^{2}_{s}}{2\beta \beta'} d\beta' \nonumber\\
\end{eqnarray}
for $\beta_{s} \leq \beta$.
When the black hole, the observer and the sun are perfectly aligned,
we obtain $s(\beta)$ as
\begin{equation}
s(0)=\frac{2}{\beta_{s}}.
\end{equation}
From Eqs.~(\ref{eq:theta+1}) and (\ref{eq:magnification1}), the magnifications $\mu_{+}(\beta)$ are given by~\cite{Bozza:2004kq}
\begin{equation}
\mu_{+}(\beta)=-\frac{D_{OS}^{2}}{D_{LS}^{2}}\frac{\theta_{m}^{2}e^{(\bar{b}-\pi)/\bar{a}}\left[ 1+e^{(\bar{b}-\pi)/\bar{a}} \right]}{\bar{a}}s(\beta).
\end{equation}

A negative solution $\theta=\theta_{-}(\beta)$ of the Ohanian lens equation is given by
\begin{equation}\label{eq:negative_image}
\theta_{-}(\beta)= -\theta_{+}(-\beta) \sim -\theta_{+}(\beta)
\end{equation}
and its magnification $\mu_{-}(\beta)$ is obtained as
\begin{equation}
\mu_{-}(\beta)
\sim-\mu_{+}(\beta)
\end{equation}
because of the spherical symmetry.
The total magnification $\mu_{tot}(\beta)$ of the double image is given by
\begin{eqnarray}
\mu_{tot}(\beta)
&\equiv& \left|  \mu_{+}(\beta) \right| +\left| \mu_{-}(\beta) \right| \nonumber\\
&=& 2\frac{D_{OS}^{2}}{D_{LS}^{2}}\frac{\theta_{m}^{2}e^{(\bar{b}-\pi)/\bar{a}}\left[ 1+e^{(\bar{b}-\pi)/\bar{a}} \right]}{\bar{a}}\left|s(\beta)\right|. \nonumber\\
\end{eqnarray}
The total magnification in a perfect-aligned case is obtained as
\begin{equation}\label{eq:aligned_magnification}
\mu_{tot}(0)=4\frac{D_{OS}^{2}}{D_{LS}^{2}}\frac{\theta_{m}^{2}e^{(\bar{b}-\pi)/\bar{a}}\left[ 1+e^{(\bar{b}-\pi)/\bar{a}} \right]}{\bar{a}\beta_{s}}.
\end{equation}

\subsection{Light curve}
\begin{figure}[htbp]
\begin{center}
\includegraphics[width=70mm]{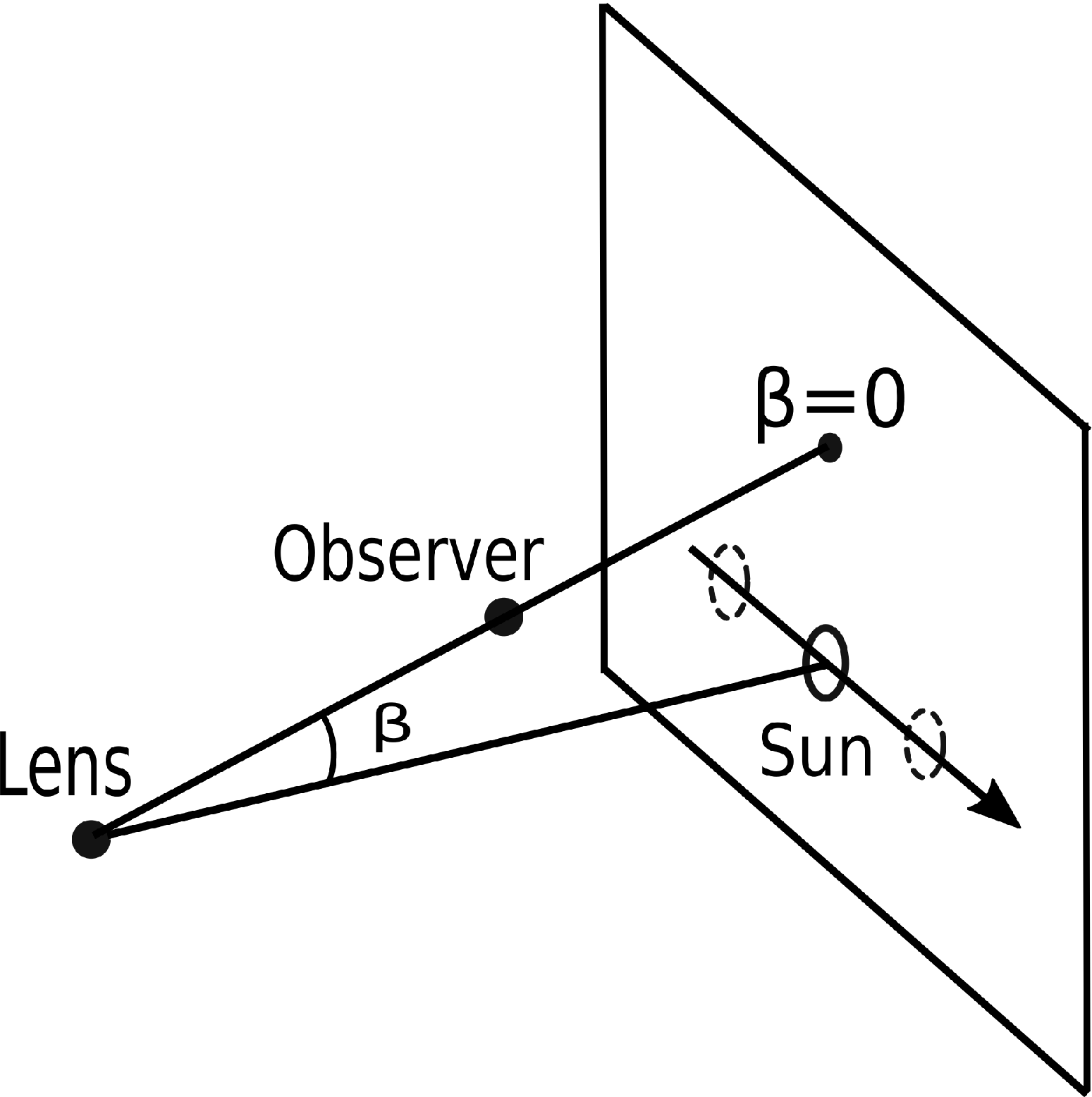}
\end{center}
\caption{The sun moves on the source plane with the orbital velocity.}
\label{microlens2}
\end{figure}

\begin{figure}[htbp]
\begin{center}
\includegraphics[width=70mm]{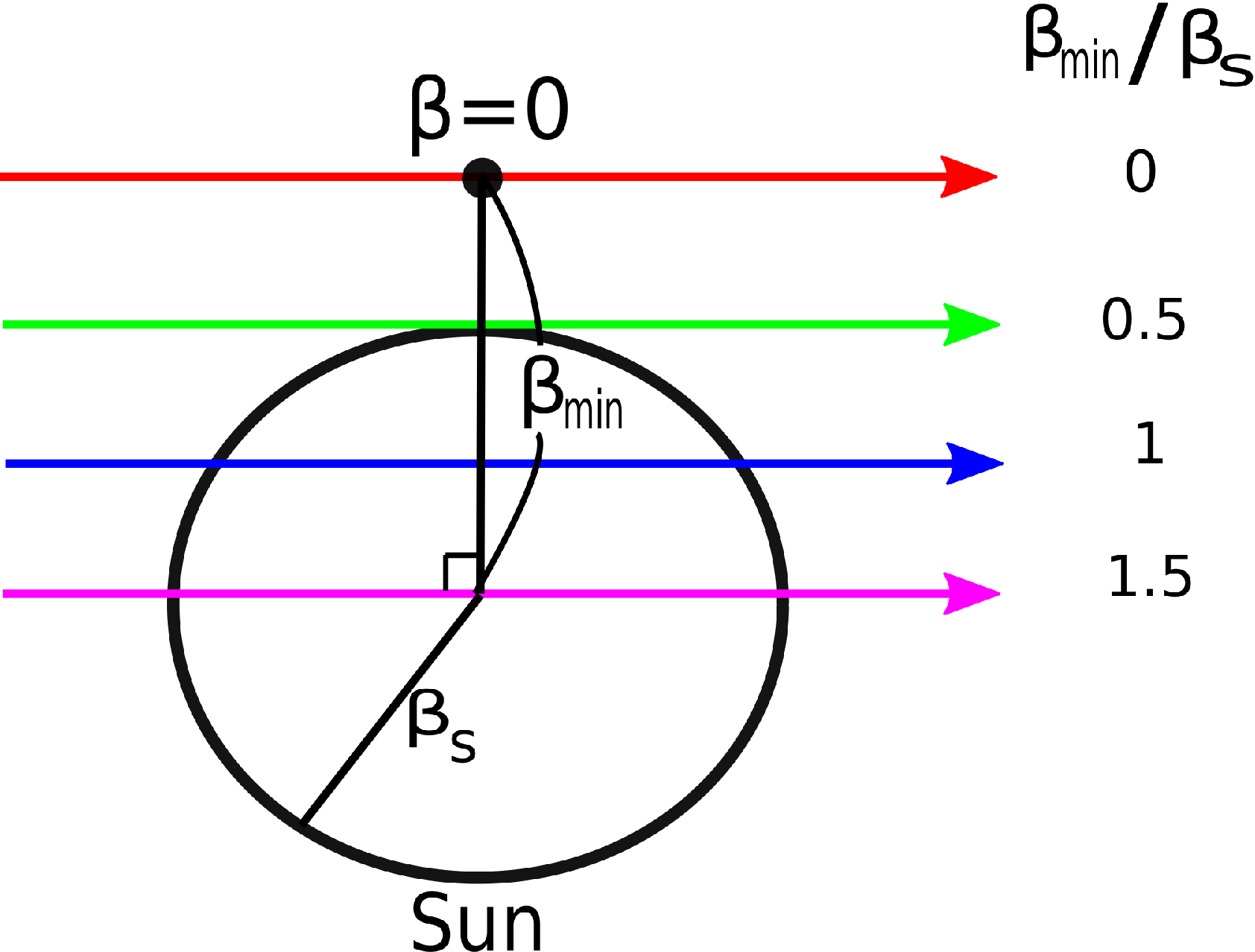}
\end{center}
\caption{The motion of the sun with the closest separation $\beta_{min}$ between the center of the sun disk and the axis $\beta=0$ on the source plane.
The (red) first, (green) second, (blue) third, and (purple) fourth right arrows from the top to bottom denote
the cases for $\beta_{min}=0$, $0.5\beta_{s}$, $\beta_{s}$, and $1.5\beta_{s}$, respectively.}
\label{microlens}
\end{figure}

\begin{figure}[htbp]
\begin{center}
\includegraphics[width=70mm]{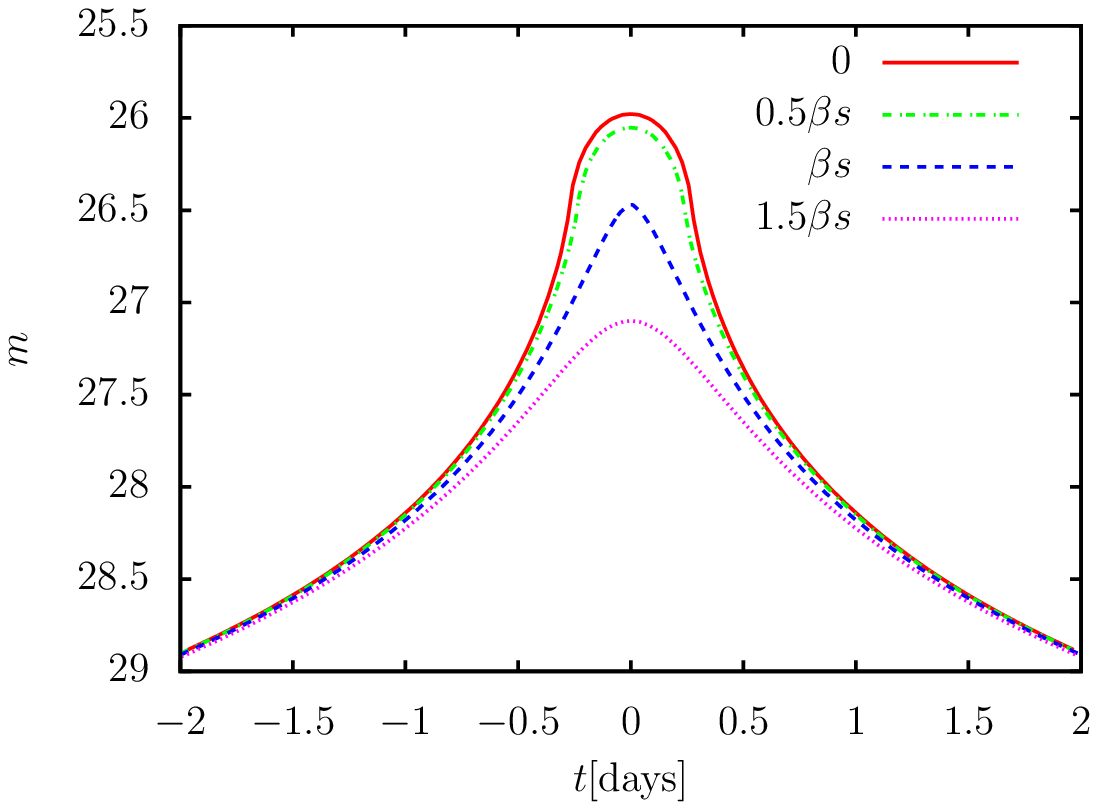}
\end{center}
\caption{Retrolensing light curves by a non-charged black hole with the mass $M=10M_{\odot}$ at $D_{OL}=0.01$ pc.
The (red) solid, (green) dash-dotted, (blue) dashed, and (purple) dotted curves denote the light curves
with the closest separation $\beta_{min}=0$, $0.5\beta_{s}$, $\beta_{s}$, and $1.5\beta_{s}$, respectively.}
\label{beta_min}
\end{figure}

We consider retrolensing light curves by a black hole at the distance $D_{OL}=0.01$pc
with the sun moving with the orbital velocity of the sun on the source plane,
as shown in Fig.~\ref{microlens2}.
We define $\beta_{min}$ as the closest separation between the center of the sun disk and the axis $\beta=0$ on the source plane,
see Fig.~\ref{microlens} for the closest separation $\beta_{min}$.
The light curves by a black hole with the mass $M=10M_{\odot}$ and with a vanishing charge $Q=0$ are shown in Fig.~\ref{beta_min}.
We consider the cases for $\beta_{min}=0$, $0.5\beta_{s}$, $\beta_{s}$, and $1.5\beta_{s}$.
The light curves show characteristic shapes at the peak depending on a reduced closest separation $\beta_{min}/\beta_{s}$.
From the shape of the peak, we would estimate the reduced closest separation $\beta_{min}/\beta_{s}$
with an accuracy which is sufficient for the following discussions.
Please note also that the light curves do not diverge in the perfect-aligned case $\beta_{min}=0$
and the peak magnitude is given by Eq.~(\ref{eq:aligned_magnification}).

\begin{figure}[htbp]
\begin{center}
\includegraphics[width=70mm]{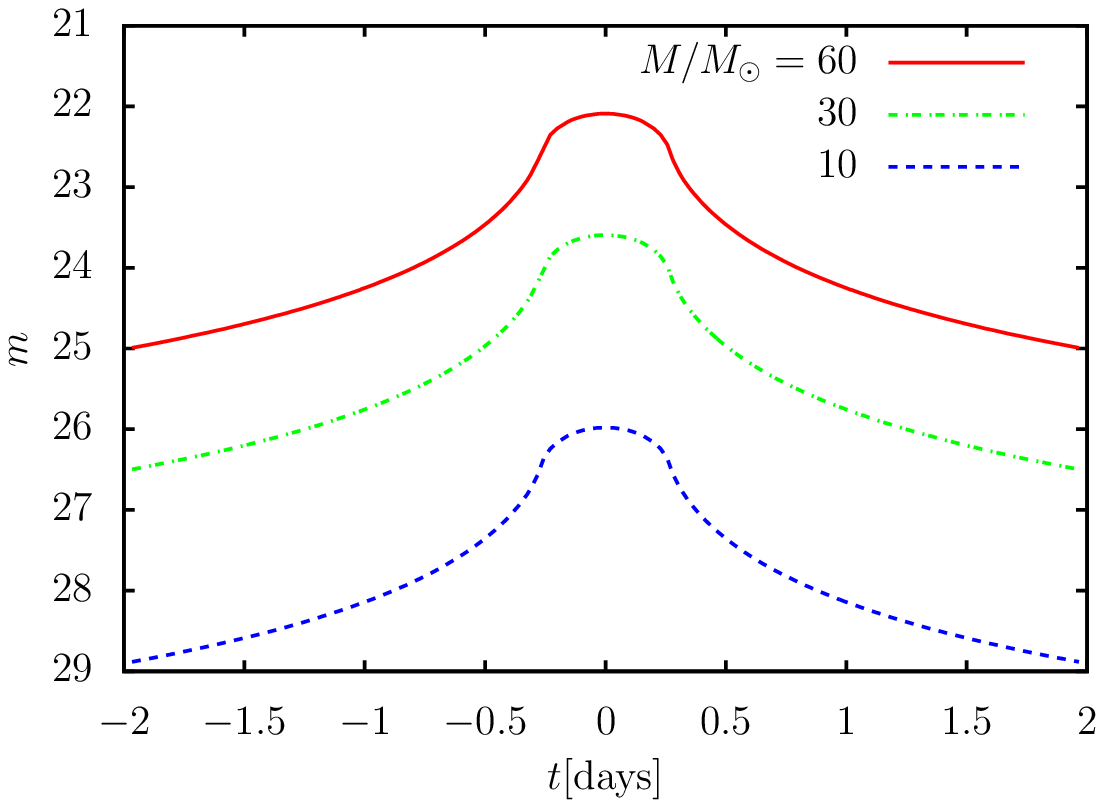}
\end{center}
\caption{Retrolensing light curves reflected by a non-charged black hole $(Q=0)$ at $D_{OL}=0.01$pc in the perfect-aligned case ($\beta_{min}=0$).
The (red) solid, (green) dash-dotted, and (blue) dashed curves denote the light curves
lensed by black holes with the mass $M=60M_{\odot}$, $30M_{\odot}$, and $10M_{\odot}$, respectively.}
\label{c0q00}
\end{figure}

\begin{figure}[htbp]
\begin{center}
\includegraphics[width=70mm]{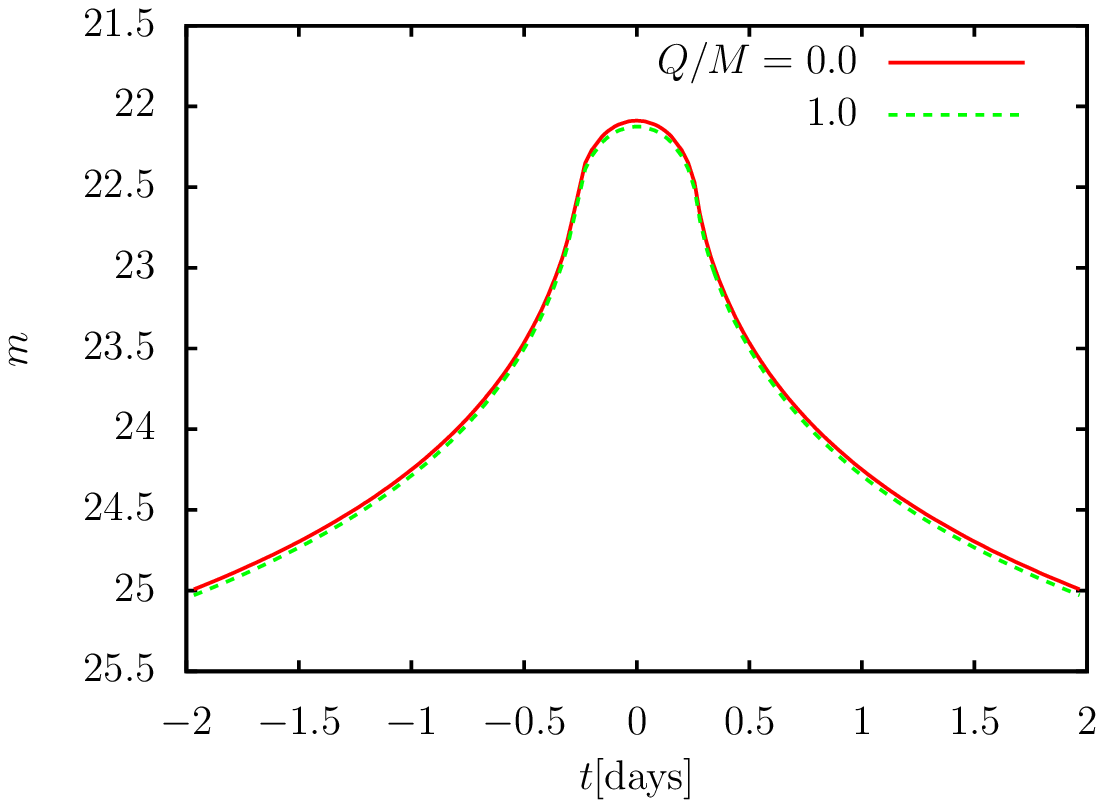}
\end{center}
\caption{Retrolensing light curves in the perfect-aligned case ($\beta_{min}=0$) lensed by a black hole with the mass $M=60M_{\odot}$ at $D_{OL}=0.01$pc.
The (red) solid and (green) dash-dotted curves denote the light curves
lensed by the non-charged black hole ($Q=0$) and the maximal charged black hole ($Q=M$), respectively.}
\label{c0m60}
\end{figure}

\begin{figure}[htbp]
\begin{center}
\includegraphics[width=70mm]{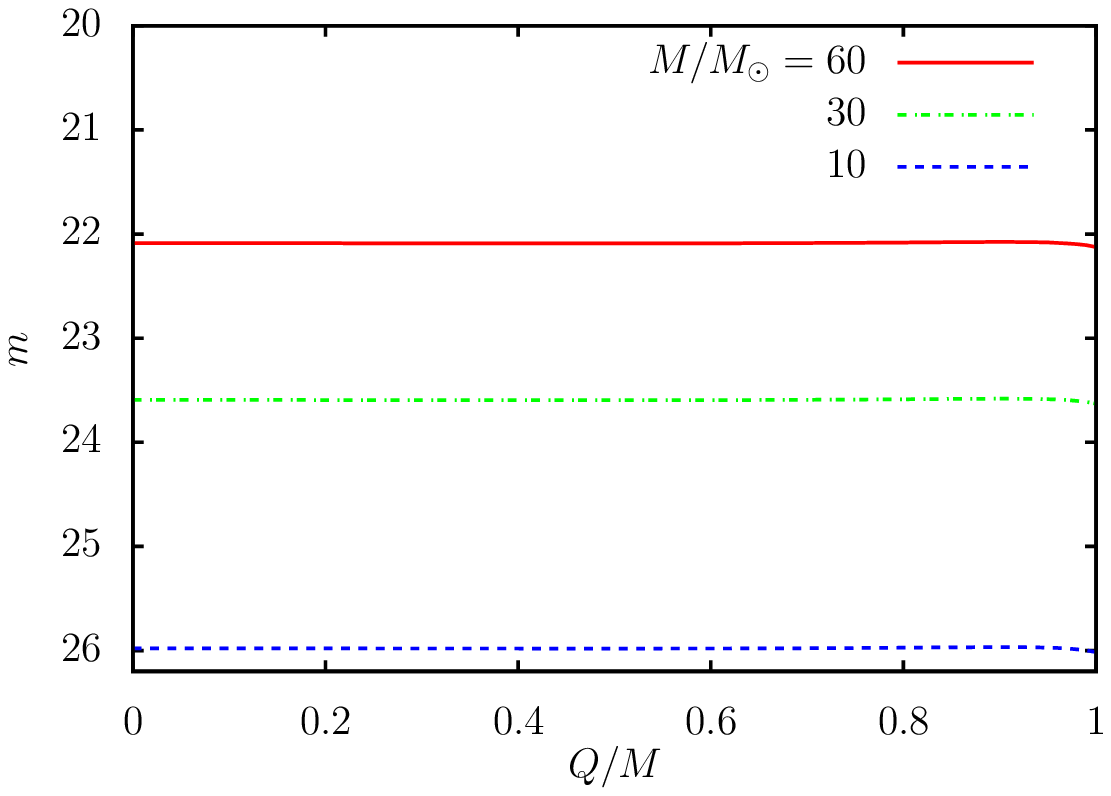}
\end{center}
\caption{The apparent magnitude $m$ of the light curves at the peak in the perfect-aligned case ($\beta_{min}=0$)
by a black hole at $D_{OL}=0.01$pc as a function of $Q/M$.
The (red) solid, (green) dash-dotted, and (blue) dashed curves denote the apparent magnitude $m$ of the peak of the light curve
lensed by black holes with the mass $M=60M_{\odot}$, $30M_{\odot}$, and $10M_{\odot}$, respectively.}
\label{pk}
\end{figure}

Figure~\ref{c0q00} shows the retrolensing light curves reflected by black holes with the mass $M=10M_{\odot}$, $30M_{\odot}$, and $60M_{\odot}$
and zero electric charge at $D_{OL}=0.01$pc in the perfect-aligned case ($\beta_{min}=0$).
The light curves show clearly that the apparent brightness of the double image or its total magnification $\mu_{tot}$ is in proportion to $M^{2}$.
Figure~\ref{c0m60} shows that two light curves by a black hole with the mass $M=60M_{\odot}$ and with the electrical charge $Q=0$ and $Q=M$
at $D_{OL}=0.01$pc in the perfect-aligned case ($\beta_{min}=0$).
We notice that the electric charge of the black hole does not change the light curve much.
The apparent magnitude $m$ of the light curves at the peak in the perfect-aligned case ($\beta_{min}=0$) as a function of $Q/M$ is shown in Fig.~\ref{pk}.

\begin{figure}[htbp]
\begin{center}
\includegraphics[width=70mm]{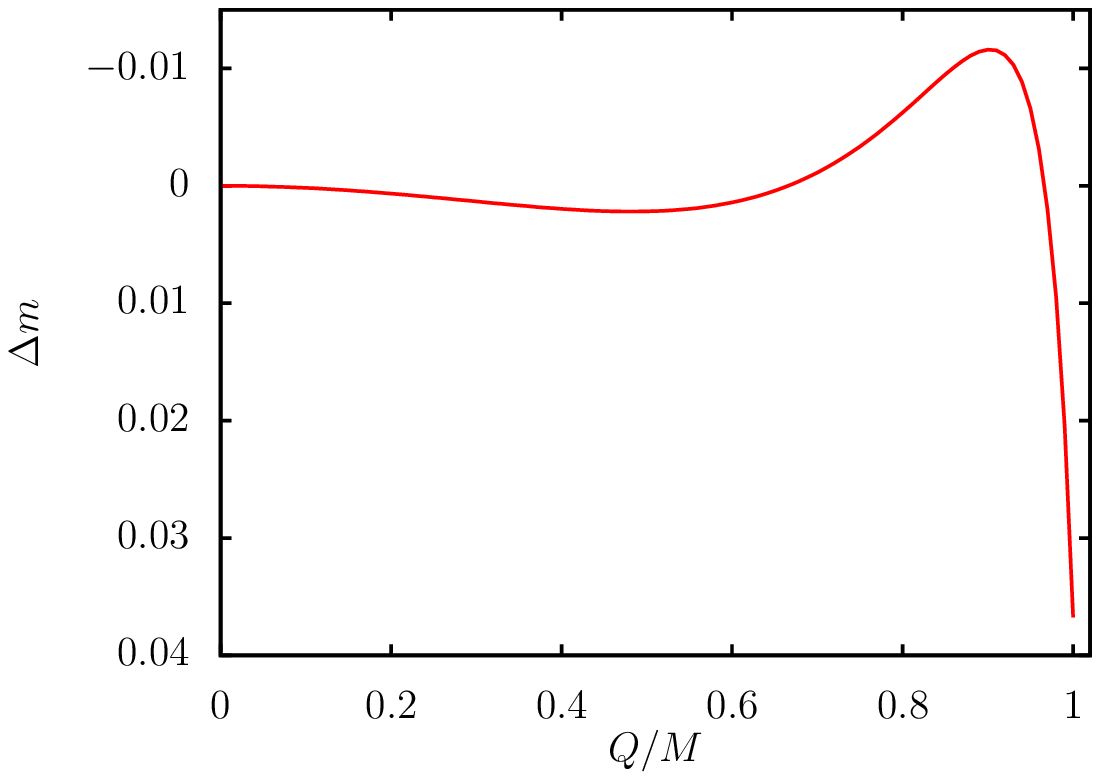}
\end{center}
\caption{The relative magnitude $\Delta m$ by a black hole at $D_{OL}=0.01$pc in the perfect-aligned case ($\beta_{min}=0$) as a function of $Q/M$.
The relative magnitude $\Delta m$ for the black hole with the mass $M=60M_{\odot}$ is plotted
but the curves with $M=10M_{\odot}$ and $M=30M_{\odot}$ are very similar to the one with $M=60M_{\odot}$.}
\label{dpk}
\end{figure}

We consider the relative magnitude $\Delta m$ of the peak of a light curve by a charged black hole
with respect to the apparent magnitude of the one by a non-charged black hole with the same mass and position.
Figure~\ref{dpk} shows the relative magnitude $\Delta m$ as a function of $Q/M$.
The relative magnitude $\Delta m$ for the black hole at $D_{OL}=0.01$pc with the mass $M=60M_{\odot}$ in the perfect-aligned case is shown in Fig.~\ref{dpk}, and
those with $M=10M_{\odot}$ and $M=30M_{\odot}$ are very similar to that with $M=60M_{\odot}$.
From
Fig.~\ref{dpk},
we see that the peak magnitude of the light curve does not change monotonically as the electric charge $Q$ increases.

\subsection{Retrolensing double image}
\begin{figure}[htbp]
\begin{center}
\includegraphics[width=70mm]{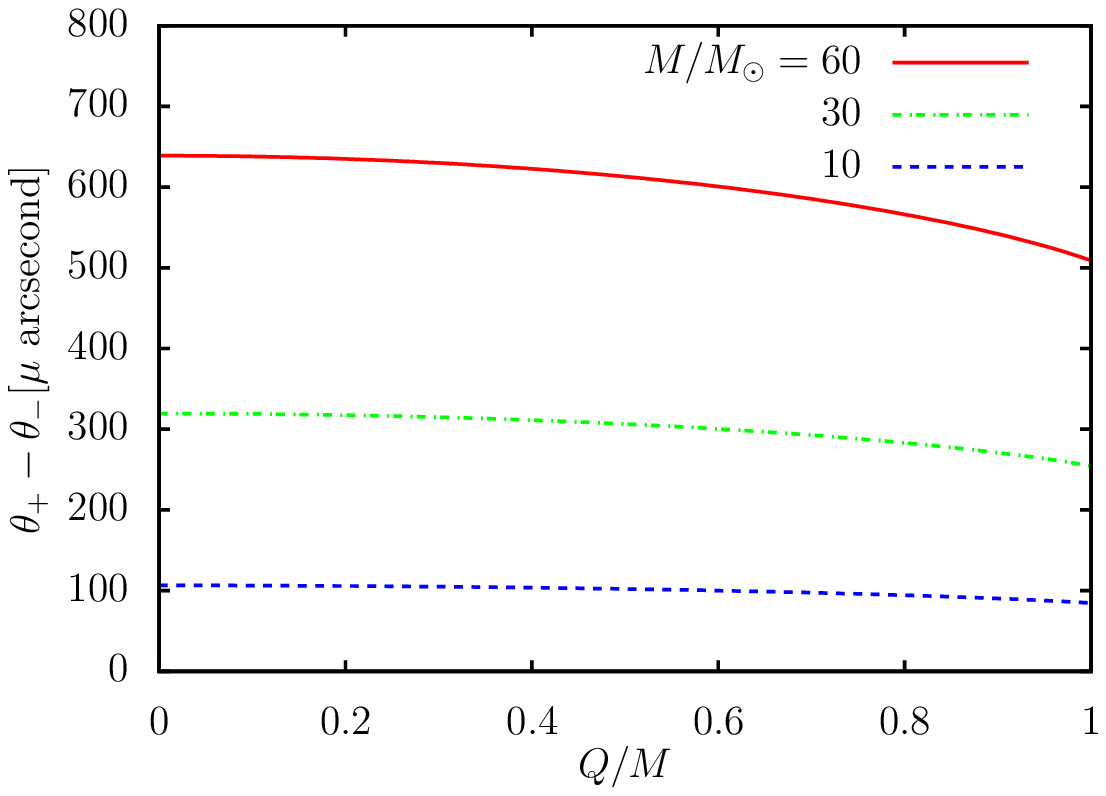}
\end{center}
\caption{The separation $\theta_{+}-\theta_{-}$ of a double image as a function of $Q/M$.
The (red) solid, (green) dash-dotted, and (blue) dashed curves denote the separation $\theta_{+}-\theta_{-}$ of a double image
lensed by black holes at $D_{OL}=0.01$pc with the mass $M=60M_{\odot}$, $30M_{\odot}$, and $10M_{\odot}$, respectively.}
\label{separation}
\end{figure}

\begin{figure}[htbp]
\begin{center}
\includegraphics[width=70mm]{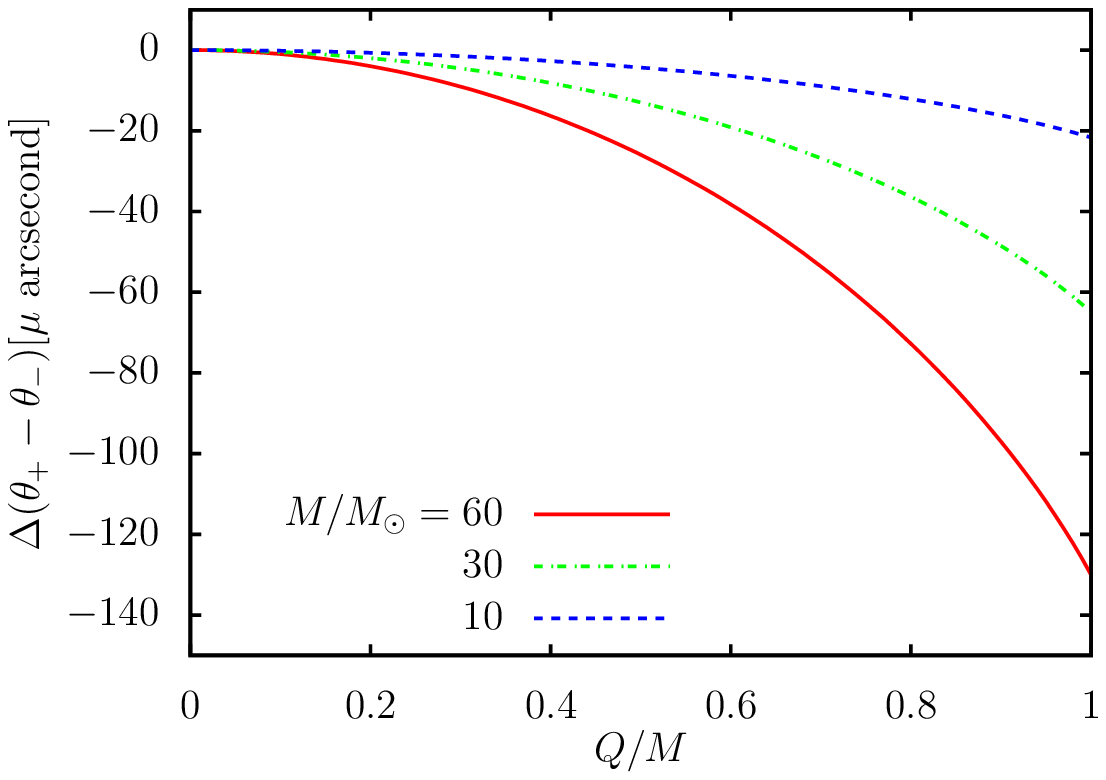}
\end{center}
\caption{The relative separation $\Delta (\theta_{+}-\theta_{-})$ as a function of $Q/M$.
The (red) solid, (green) dash-dotted, and (blue) dashed curves denote the relative separation $\Delta (\theta_{+}-\theta_{-})$
by black holes at $D_{OL}=0.01$pc with the masses $M=60M_{\odot}$, $30M_{\odot}$, and $10M_{\odot}$, respectively.}
\label{dseparation}
\end{figure}

We discuss a retrolensing double image with image angles $\theta_{+}$ and $\theta_{-}$ by a charged black hole at $D_{OL}=0.01$pc.
From Eqs.~(\ref{eq:theta+1}) and (\ref{eq:negative_image}), the separation $\theta_{+}-\theta_{-}$ of the double image is given by
\begin{equation}\label{eq:separation1}
\theta_{+}-\theta_{-}
\sim 2\theta_{+}
=2\theta_{m} \left[ 1+\exp\left( \frac{\bar{b}-\pi+\beta}{\bar{a}} \right) \right].
\end{equation}
We define the relative separation $\Delta (\theta_{+}-\theta_{-})$
as the difference of the separation $\theta_{+}-\theta_{-}$ of the double image retrolensed by the charged black hole
from the one by the non-charged black hole.
Figures~\ref{separation} and \ref{dseparation} show the separation $\theta_{+}-\theta_{-}$ of the double image
and the relative separation $\Delta (\theta_{+}-\theta_{-})$, respectively.

\section{Discussion and Conclusion}
In this paper, in the strong deflection limit we obtained the deflection angle analytically in the Reissner-Nordstr\"{o}m spacetime
while it was not obtained in Ref.~\cite{Bozza_2002}.
This is because we chose the variable $z$ defined in Eq.~(\ref{eq:z1}).
In Ref.~\cite{Bozza_2002} a variable $z_{[17]}$ defined by, in our notation,
\begin{equation}
z_{[17]}\equiv \frac{-g_{tt}(r)+g_{tt}(r_{0})}{1+g_{tt}(r_{0})}
\end{equation}
was used instead,
see Eqs. (10) and (11) in~\cite{Bozza_2002}.
Since $z_{[17]}$ in the Reissner-Nordstr\"{o}m spacetime becomes
\begin{equation}
z_{[17]}=1-\frac{r_{0}^{2}\left(2Mr-Q^{2}\right)}{r^{2}\left(2Mr_{0}-Q^{2}\right)},
\end{equation}
$z_{[17]}$ is not equivalent to $z$ defined by Eq.~(\ref{eq:z1}).
We also notice that $z_{[17]}$ is the same as $z$ when the charge $Q$ vanishes.
Using the variable $z$ defined in Eq.~(\ref{eq:z1}),
the deflection angle in the strong deflection limit in a general asymptotically flat, static, spherically symmetric spacetime will be developed in a
follow-up publication~\cite{Tsukamoto:2016jzh}.

Eiroa and Torres mainly discussed the image separations and the magnifications of images retrolensed by
a supermassive black hole in Galactic center and a stellar mass black hole in the Galactic hole in Ref.~\cite{Eiroa:2003jf}.
In this paper, using the obtained deflection angle in the strong deflection limit in the Reissner-Nordstr\"{o}m spacetime,
we considered the details of light curves and the double images retrolensed by a stellar mass black hole near the solar system
which might be measured in the near future.
We have shown that the effect of the electric charge $Q$ of the black hole on the magnitude of the light curve is small
even if the black hole has the maximal electric charge $Q=M$.
The magnitude of the peak of the the light curve depends on the reduced closest separation $\beta_{min}/\beta_{s}$
between the center of the sun and the axis $\beta=0$ on the source plane
and one would estimate the reduced closest separation $\beta_{min}/\beta_{s}$ from the characteristic shape of the peak of the light curve.
Thus, unknown parameters of retrolensing are the distance $D_{OL}$ between the observer and the black hole,
the mass $M$ and the electric charge $Q$ of the black hole.

We also considered the separation of the double images which appear near the outside of the light sphere of the charged black hole.
If the black hole has the maximal charge $Q=M$,
the image separation is $20$ percent smaller than the one by a non-charged black hole with the same mass.
The image separations of retrolensing by black holes with the mass $M=10M_{\odot}$, $30M_{\odot}$, and $60M_{\odot}$
at $D_{OL}=0.03$pc, $0.1$pc, and  $0.2$pc,
respectively, are $30 \mu$arcsecond, which are the same size of the separation of the double image of light rays reflected by
the light sphere of the black hole at the center of our galaxy~\cite{Virbhadra_Ellis_2000,Bozza_2002}.
If we are lucky, we may measure the image separation near outside the light sphere of a black hole passing by the solar system
in the near future.

If one believes that a non-rotating black hole should be described well by a Schwarzschild black hole,
one can determine the mass $M$ and its distance $D_{OL}$ from the measurement of the image separation and the peak magnitude of the light curve.
We have concentrated on retrolensing but if one can measure another observable in a weak or strong gravitational field or the details of retrolensing,
one can determine an additional parameter like $Q$ or one can confirm that the lens object can be really regarded as the Schwarzschild black hole.

If one observes a light curve with precisely solar spectra on the ecliptic,
one can say that it will be retrolensing caused by a light sphere as Holz and Wheeler pointed out~\cite{Holz:2002uf}.
The retrolens can be a black hole, a wormhole, or other dark and compact objects with a light sphere
and the observer will not distinguish them by the shapes of their retrolensing light curves
since they will be very similar with each other~\cite{Tsukamoto:2016zdu}.
Even if one cannot detect any retrolensing light curves,
one can give an upper bound of the number density of dark and compact objects with a light sphere from the observational data.
In this paper, we have only considered a Reissner-Nordstr\"{o}m black hole as a simple example
but our conclusion would not change much in the other non-rotating black hole spacetimes which include the Schwarzschild black hole spacetime as a special case.
The consideration of the additional measurement of a black hole passing by the solar system without retrolensing
and the further consideration of retrolensing by a Reissner-Nordstr\"{o}m black hole
and other non-rotating black holes we leave for future work.

\section*{Acknowledgements}
N.~T. thanks Ken-ichi Nakao, Tetsuya Shiromizu, Chul-Moon Yoo, and Takahisa Igata for valuable comments.
He also thanks Tomohiro Harada, Yoshimune Tomikawa, Hideki Asada, Hirotaka Yoshino, Yusuke Suzuki, Rio Saitou, Masato Nozawa,
and Takafumi Kokubu for useful conversations.
The authors are grateful to Carlos~A.~R.~Herdeiro for bringing their attention to gravitational lensing by boson stars.
This research was supported in part by the National Natural Science Foundation of China under Grant No. 11475065,
the Major Program of the National Natural Science Foundation of China under Grant No. 11690021.
%


\begin{thebibliography}{99}



\bibitem{Schneider_Ehlers_Falco_1992}
P. Schneider, J. Ehlers, and E. E. Falco,
\textit{Gravitational Lenses} (Springer-Verlag, Berlin, 1992).

\bibitem{Petters_Levine_Wambsganss_2001}
A. O. Petters, H. Levine, and J. Wambsganss,
\textit{Singularity Theory and Gravitational Lensing} (Birkhauser, Boston, 2001).

\bibitem{Schneider_Kochanek_Wambsganss_2006}
P. Schneider, C. S. Kochanek, and J. Wambsganss,
\textit{Gravitational Lensing: Strong, Weak and Micro,
Lecture Notes of the 33rd Saas-Fee Advanced Course},
edited by G. Meylan, P. Jetzer, and P. North (Springer-Verlag, Berlin, 2006).

\bibitem{Bartelmann_2010}
M.~Bartelmann,
Class.\ Quant.\ Grav.\  {\bf 27}, 233001 (2010).

\bibitem{Perlick_2004_Living_Rev}
V. Perlick,
Living Rev. Relativity {\bf7}, 9 (2004).

\bibitem{Bozza_2010}
V. Bozza,
Gen. Relativ. Gravit. {\bf 42}, 2269 (2010).

\bibitem{Claudel:2000yi}
  C.~M.~Claudel, K.~S.~Virbhadra, and G.~F.~R.~Ellis,
  J.\ Math.\ Phys.\  {\bf 42}, 818 (2001).

\bibitem{Hasse_Perlick_2002}
W. Hasse and V. Perlick,
Gen. Relativ. Gravit. {\bf 34}, 415 (2002).

\bibitem{Darwin_1959}
C. Darwin,
Proc. R. Soc. Lond. A {\bf 249}, 180 (1959).

\bibitem{Atkinson_1965}
R.~d'~E. Atkinson,
Astron. J. {\bf 70}, 517 (1965).

\bibitem{Luminet_1979}
J.-P. Luminet,  Astron. Astrophys. {\bf 75}, 228 (1979).

\bibitem{Ohanian_1987}
H. C. Ohanian, Am. J. Phys. {\bf 55}, 428 (1987).

\bibitem{Nemiroff_1993}
R. J. Nemiroff,  Am. J. Phys. {\bf 61}, 619 (1993).

\bibitem{Frittelli_Kling_Newman_2000}
S. Frittelli, T. P. Kling, and E. T. Newman,
Phys. Rev. D {\bf 61}, 064021 (2000).

\bibitem{Virbhadra_Ellis_2000}
K. S. Virbhadra and G. F. R. Ellis,
Phys. Rev. D {\bf 62}, 084003 (2000).

\bibitem{Bozza_Capozziello_Iovane_Scarpetta_2001}
V. Bozza, S. Capozziello, G. Iovane, and G. Scarpetta,
Gen. Relativ. Gravit. {\bf 33}, 1535 (2001).

\bibitem{Bozza_2002}
V. Bozza,
Phys. Rev. D {\bf 66}, 103001 (2002).

\bibitem{Perlick_2004_Phys_Rev_D}
V. Perlick,
Phys. Rev. D {\bf 69}, 064017 (2004).

\bibitem{Iyer:2006cn}
  S.~V.~Iyer and A.~O.~Petters,
  Gen.\ Rel.\ Grav.\  {\bf 39}, 1563 (2007).

\bibitem{Bozza_Sereno_2006}
V. Bozza and M. Sereno,
Phys. Rev. D {\bf 73}, 103004 (2006).

\bibitem{Bozza:2007gt}
V.~Bozza and G.~Scarpetta,
Phys.\ Rev.\ D {\bf 76}, 083008 (2007).

\bibitem{Bozza:2008ev}
V.~Bozza,
Phys.\ Rev.\ D {\bf 78}, 103005 (2008).

\bibitem{Virbhadra_2009}
K. S. Virbhadra,
Phys. Rev. D {\bf 79}, 083004 (2009).

\bibitem{Ishihara:2016sfv}
  A.~Ishihara, Y.~Suzuki, T.~Ono and H.~Asada,
  Phys.\ Rev.\ D {\bf 95}, 044017 (2017).

\bibitem{Bozza:2002af}
  V.~Bozza,
  Phys.\ Rev.\ D {\bf 67}, 103006 (2003).

\bibitem{Eiroa:2002mk}
  E.~F.~Eiroa, G.~E.~Romero, and D.~F.~Torres,
  Phys.\ Rev.\ D {\bf 66}, 024010 (2002).

\bibitem{Bozza:2006nm}
  V.~Bozza, F.~De Luca, and G.~Scarpetta,
  Phys.\ Rev.\ D {\bf 74}, 063001 (2006).

\bibitem{Bozza:2005tg}
  V.~Bozza, F.~De Luca, G.~Scarpetta, and M.~Sereno,
  Phys.\ Rev.\ D {\bf 72}, 083003 (2005).

\bibitem{Saida:2016kpk}
  H.~Saida,
  arXiv:1606.04716 [astro-ph.HE].

\bibitem{Chen:2016hil}
  S.~Chen, S.~Wang, Y.~Huang, J.~Jing, and S.~Wang,
  arXiv:1611.08783 [gr-qc].

\bibitem{Eiroa:2004gh}
  E.~F.~Eiroa,
  Phys.\ Rev.\ D {\bf 71}, 083010 (2005).

\bibitem{Whisker:2004gq}
  R.~Whisker,
  Phys.\ Rev.\ D {\bf 71}, 064004 (2005).

\bibitem{Eiroa:2012fb}
  E.~F.~Eiroa and C.~M.~Sendra,
  Phys.\ Rev.\ D {\bf 86}, 083009 (2012).

\bibitem{Bhadra:2003zs}
  A.~Bhadra,
  Phys.\ Rev.\ D {\bf 67}, 103009 (2003).

\bibitem{Eiroa:2005ag}
  E.~F.~Eiroa,
  Phys.\ Rev.\ D {\bf 73}, 043002 (2006).

\bibitem{Mukherjee:2006ru}
  N.~Mukherjee and A.~S.~Majumdar,
  Gen.\ Rel.\ Grav.\  {\bf 39}, 583 (2007).

\bibitem{Gyulchev:2006zg}
  G.~N.~Gyulchev and S.~S.~Yazadjiev,
  Phys.\ Rev.\ D {\bf 75}, 023006 (2007).

\bibitem{Chen:2009eu}
  S.~b.~Chen and J.~l.~Jing,
  Phys.\ Rev.\ D {\bf 80}, 024036 (2009).

\bibitem{Liu:2010wh}
  Y.~Liu, S.~Chen, and J.~Jing,
  Phys.\ Rev.\ D {\bf 81}, 124017 (2010).

\bibitem{Eiroa:2010wm}
  E.~F.~Eiroa and C.~M.~Sendra,
  Class.\ Quant.\ Grav.\  {\bf 28}, 085008 (2011).

\bibitem{Ding:2010dc}
  C.~Ding, S.~Kang, C.~Y.~Chen, S.~Chen, and J.~Jing,
  Phys.\ Rev.\ D {\bf 83}, 084005 (2011).

\bibitem{Chen:2011ef}
  S.~Chen, Y.~Liu, and J.~Jing,
  Phys.\ Rev.\ D {\bf 83}, 124019 (2011).

\bibitem{Wei:2011nj}
  S.~W.~Wei, Y.~X.~Liu, C.~E.~Fu, and K.~Yang,
  JCAP {\bf 1210}, 053 (2012).

\bibitem{AzregAinou:2012xv}
  M.~Azreg-Ainou,
  Phys.\ Rev.\ D {\bf 87}, 024012 (2013).

\bibitem{Gyulchev:2012ty}
  G.~N.~Gyulchev and I.~Z.~Stefanov,
  Phys.\ Rev.\ D {\bf 87}, 063005 (2013).

\bibitem{Eiroa:2013nra}
  E.~F.~Eiroa and C.~M.~Sendra,
  Phys.\ Rev.\ D {\bf 88}, 103007 (2013).

\bibitem{Wei:2014dka}
  S.~W.~Wei, K.~Yang, and Y.~X.~Liu,
  Eur.\ Phys.\ J.\ C {\bf 75}, 253 (2015)
  Erratum: [Eur.\ Phys.\ J.\ C {\bf 75}, 331 (2015)].

\bibitem{Eiroa:2014mca}
  E.~F.~Eiroa and C.~M.~Sendra,
  Eur.\ Phys.\ J.\ C {\bf 74}, 3171 (2014).

\bibitem{Sahu:2015dea}
  S.~Sahu, K.~Lochan, and D.~Narasimha,
  Phys.\ Rev.\ D {\bf 91}, 063001 (2015).

\bibitem{Sotani:2015ewa}
  H.~Sotani and U.~Miyamoto,
  Phys.\ Rev.\ D {\bf 92}, 044052 (2015).

\bibitem{Zhao:2016kft}
  S.~S.~Zhao and Y.~Xie,
  JCAP {\bf 1607}, 007 (2016).

\bibitem{Tsukamoto:2014dta}
  N.~Tsukamoto, T.~Kitamura, K.~Nakajima, and H.~Asada,
  Phys.\ Rev.\ D {\bf 90}, 064043 (2014).

\bibitem{Chakraborty:2016lxo}
  S.~Chakraborty and S.~SenGupta,
  arXiv:1611.06936 [gr-qc].

\bibitem{Virbhadra_Keeton_2008}
K. S. Virbhadra and C. R. Keeton,
Phys. Rev. D {\bf 77}, 124014 (2008).

\bibitem{Virbhadra_Ellis_2002}
K. S. Virbhadra and G. F. R. Ellis,
Phys. Rev. D {\bf 65}, 103004 (2002).

\bibitem{Dey_Sen_2008}
T. K. Dey and S. Sen,
Mod. Phys. Lett. A {\bf 23}, 953 (2008).

\bibitem{Horvat:2013plm}
  D.~Horvat, S.~Ilijic, A.~Kirin, and Z.~Narancic,
  Class.\ Quant.\ Grav.\  {\bf 30}, 095014 (2013).

\bibitem{Cunha:2015yba}
  P.~V.~P.~Cunha, C.~A.~R.~Herdeiro, E.~Radu, and H.~F.~Runarsson,
  Phys.\ Rev.\ Lett.\  {\bf 115}, 211102 (2015).

\bibitem{Cunha:2016bjh}
  P.~V.~P.~Cunha, J.~Grover, C.~Herdeiro, E.~Radu, H.~Runarsson, and A.~Wittig,
  Phys.\ Rev.\ D {\bf 94}, 104023 (2016).

\bibitem{Nandi_Zhang_Zakharov_2006}
K. K. Nandi, Y. Z. Zhang, and A. V. Zakharov,
Phys. Rev. D {\bf 74}, 024020 (2006).

\bibitem{Tsukamoto_Harada_Yajima_2012}
N. Tsukamoto, T. Harada, and K. Yajima,
Phys. Rev. D {\bf 86}, 104062 (2012).

\bibitem{Nandi:2016ccg}
  K.~K.~Nandi, A.~A.~Potapov, R.~N.~Izmailov, A.~Tamang, and J.~C.~Evans,
  Phys.\ Rev.\ D {\bf 93}, no. 10, 104044 (2016).

\bibitem{Tsukamoto:2016qro}
  N.~Tsukamoto,
  Phys.\ Rev.\ D {\bf 94}, 124001 (2016).

\bibitem{Tsukamoto:2016zdu}
  N.~Tsukamoto and T.~Harada,
  Phys.\ Rev.\ D {\bf 95}, 024030 (2017).

\bibitem{Abbott:2016blz}
  B.~P.~Abbott {\it et al.} [LIGO Scientific and Virgo Collaborations],
  Phys.\ Rev.\ Lett.\  {\bf 116}, 061102 (2016).

\bibitem{TheLIGOScientific:2016htt}
  B.~P.~Abbott {\it et al.} [LIGO Scientific and Virgo Collaborations],
  Astrophys.\ J.\  {\bf 818}, L22 (2016).

\bibitem{Holz:2002uf}
  D.~E.~Holz and J.~A.~Wheeler,
  Astrophys.\ J.\  {\bf 578}, 330 (2002).

\bibitem{Eiroa:2003jf}
  E.~F.~Eiroa and D.~F.~Torres,
  Phys.\ Rev.\ D {\bf 69}, 063004 (2004).

\bibitem{Bozza:2004kq}
  V.~Bozza and L.~Mancini,
  Astrophys.\ J.\  {\bf 611}, 1045 (2004)

\bibitem{DePaolis:2003ad}
  F.~De Paolis, G.~Ingrosso, A.~Geralico, and A.~A.~Nucita,
  Astron.\ Astrophys.\  {\bf 409}, 809 (2003).

\bibitem{DePaolis:2004xe}
  F.~De Paolis, A.~Geralico, G.~Ingrosso, A.~A.~Nucita, and A.~Qadir,
  Astron.\ Astrophys.\  {\bf 415}, 1 (2004).

\bibitem{Stefanov:2010xz}
  I.~Z.~Stefanov, S.~S.~Yazadjiev, and G.~G.~Gyulchev,
  Phys.\ Rev.\ Lett.\  {\bf 104}, 251103 (2010).

\bibitem{Wei:2013mda}
  S.~W.~Wei and Y.~X.~Liu,
  Phys.\ Rev.\ D {\bf 89}, 047502 (2014).

\bibitem{Wei:2011zw}
  S.~W.~Wei, Y.~X.~Liu, and H.~Guo,
  Phys.\ Rev.\ D {\bf 84}, 041501 (2011).

\bibitem{Witt:1994}
H.~J.~Witt and S.~Mao, ApJ, {\bf 430}, 505 (1994).

\bibitem{Nemiroff:1994uz}
  R.~J.~Nemiroff and W.~A.~D.~T.~Wickramasinghe,
  Astrophys.\ J.\  {\bf 424}, L21 (1994).

\bibitem{Alcock:1997fi}
  C.~Alcock {\it et al.} [MACHO and GMAN Collaborations],
  Astrophys.\ J.\  {\bf 491}, 436 (1997).

\bibitem{Tsukamoto:2016jzh}
  N.~Tsukamoto,
  arXiv:1612.08251 [gr-qc].


\end{thebibliography}
\end{document}